# Electron diffusivities in MgB$_2$ from point contact spectroscopy

Y. Bugoslavsky, Y. Miyoshi, G.K. Perkins, A.D. Caplin, L.F. Cohen

Blackett Laboratory, Imperial College London, SW7 2BZ, United Kingdom

A. V. Pogrebnyakov, X. X. Xi

Department of Physics and Materials Research Institute, Pennsylvania State University, University Park, PA 16802

We demonstrate that the variation of the Andreev reflection with applied magnetic field provides a direct means of comparing the properties of MgB$_2$ with the theory for a dirty two-band superconductor, and we find good agreement between the two. The ratio of electron diffusivities in the σ and π bands can be inferred from this experiment. We find that the field dependence of the density of states at the Fermi level in the π band is independent of the field direction, and that the anisotropic upper critical field is determined by the anisotropic diffusivity in the σ band.

Magnesium diboride, MgB$_2$ is as yet a unique superconductor that has two distinct order parameters. They are derived from two groups of disconnected sheets of the Fermi surface (FS). Quasi two-dimensional $\sigma$ bands give rise to the larger superconducting gap, $\Delta_\sigma \sim 7$ meV, whereas the smaller gap, $\Delta_\pi$ opens up on the three-dimensional sheets of the FS that correspond to the $\pi$ bands [1,2]. First principle calculations of the electronic structure were successful in explaining the key normal-state properties of MgB$_2$. The theoretical description of superconducting properties in the presence of two bands has been developed on the basis of the Usadel equations, which are applicable to dirty-limit superconductors. Using this approach, the temperature dependence of the upper critical field $H_{c2}$ [3], the variation of the order parameters at the vortex core [4], the anisotropy of $H_{c2}$ and the London penetration depth [3,5] have all been calculated. The theoretical results are generally in good agreement with the available experiments, and provide guidance to engineering the material with desired properties. The elastic scattering rates in the two types of bands can be very different, as they are controlled by different kinds of atomic defects [6]. Varying the ratio of electron diffusivities in the two bands ($D_\sigma/D_\pi$) will change the resulting macroscopic superconducting properties, and therefore independent determination of this ratio has emerged as a key prerequisite to understand and control the properties of MgB$_2$. Due to the crystalline anisotropy, the diffusivities along the c axis ($D_{\sigma,\pi}^{(c)}$) can be different from the in-plane values ($D_{\sigma,\pi}^{(ab)}$). It is in principle possible to obtain the scattering rates from the temperature dependence of the resistivity [7,8], but this method is subject to large uncertainty if the sample is not 100% dense [9], which is often the case with MgB$_2$. In this work we demonstrate that the diffusivity ratio can be directly evaluated from point-contact Andreev reflection (PCAR) spectroscopy in magnetic field.

Point-contact [10,11,12] and tunnelling spectroscopies [13,14], have been used to study the properties of MgB$_2$ in applied magnetic field. The scanning tunneling spectroscopy (STS) images of vortices parallel to the c axis [13] have been successfully explained by theory [4]. Importantly, vortex imaging is demanding experimentally and probes mainly the properties of the π band. It is possible to use the STS method to probe the in-plane vortices in sufficiently thick single crystals [14], but certainly not in thin films. The effect of vortices on the point-contact measurements has not been considered.

In our previous work [15] we have demonstrated that as the field is increased, the transition to the normal state occurs simultaneously in the two superconducting sub-systems, i.e., the $\sigma$ and $\pi$ bands are characterised by a unique H$_{c2}$. Later we have studied the evolution of the PCAR spectra with magnetic field in niobium [16]; we showed that the mechanism behind the field variation of the Andreev reflection in Type-II superconductor is the presence of normal vortex cores, which leads to effective broadening of the spectra and increasing the zero-energy density of states (DOS), $N^0$ (corresponding to progressive increase of normal-state-like core excitations with increasing density of vortices). Therefore the PCAR data can be treated as a quantitative probe of the magnitude of $N^0$. The theory [4] predicts that in $MgB_2$ the DOS in the $\sigma$ and $\pi$ bands have distinctly different field dependences, which is a very specific conclusion that can be tested directly by experiment.

In this Letter we discuss our results in the light of this theoretical prediction. We have found that in field parallel to the c axis there is a good agreement with the theory, in that the $\pi$-band DOS, $N_\pi^{\,0}$, increases with field much faster than the $\sigma$-band DOS. From the experiment we can directly extract the dependence $N_\pi^{\,0}(H)$ and hence, on the basis of [4], the value of $D_\sigma/D_\pi$. The most striking result is that, in spite of the significant anisotropy of the upper critical field $H_{c2}$, there is *no difference* between the response of the $\pi$-band DOS to low fields applied parallel to the *c* axis (low $H_{c2}$), and to fields parallel to the *ab* plane (high $H_{c2}$). This observation suggests that the overall material anisotropy is primarily dictated by the very anisotropic nature of the $\sigma$ bands.

The measurements were performed on an epitaxial $MgB_2$ film grown by hybrid physical chemical vapour deposition (HPCVD) [17]. The critical temperature of the film was 39 K, an indication of negligible inter-band scattering. The value of $H_{c2} \parallel c$ was 6.5 T at 4.2 K, as obtained from resistivity measurements. The two-fold enhancement of $H_{c2}$ compared to clean single crystals suggests that the film is closer to the dirty-limit superconductivity. The anisotropy of $H_{c2}$ is a slowly varying function of temperature, increasing from 3 at T=38.5 K to 4 at 28 K. At lower temperatures the in-plane $H_{c2}$ was too high to be measured with our 8-Tesla magnet. Point contacts were made by pressing a sharpened gold tip to the surface of the film. The typical size of the contact footprint was 30 to 50 microns (see Fig. 1). The relevant length scale in a Type-II superconductor subjected to magnetic field is the distance between vortices, $a = \sqrt{\dfrac{2\Phi_0}{\sqrt{3}H}}$, where $\Phi_0$ is the flux quantum. For example, in a field of 1 mT, the vortex separation is about 1 $\mu$m. Since our measurements were done at significantly higher fields, there was always a large number of vortices within the contact. We therefore presume that the current from the tip samples a large area compared to the unit cell of the vortex lattice, and the results we obtain represent an effective average over the vortex lattice. We have demonstrated [16] that this interpretation provides a consistent agreement between experiment and the relevant theoretical calculations [18].

Similar to the case of Nb, here we write the conductance of the point contact in field as a sum of the "normal channel" (current injected into the cores) and the "SC channel" (current in between the vortices), extending this approach to combine it with the formulae used to describe the two-gap SC (*e.g.*, [11]). In zero field, the voltage-dependent normalised conductance of the point contact is then:

$$\frac{G(V)}{G_N} = f g_\pi + (1-f) g_\sigma. \qquad (1)$$

Here $G_N$ is the normal-state junction conductance, and the normalised conductances $g_\pi$ and $g_\sigma$ are given by the usual expressions from the Blonder-Tinkham-Klapwijk (BTK) theory [19] for the two values of the gap, $\Delta_\pi$ and $\Delta_\sigma$, respectively. The shape of the functions $g_\pi$ and $g_\sigma$ is determined by the effective interface barrier Z; we assume it is the same in the two bands. The effects of finite temperature and interface scattering are taken into account by convolution of the zero-temperature BTK functions with a Gaussian of width $\omega$. The relative weight *f* is determined by the contribution of each band to the total density of states, and by the geometry of the current injection into the SC. Unlike tunneling, direct point contacts are non-directional probes, *i.e.*, the current flows into a wide cone, effectively probing a large solid angle in the momentum space [20]. In theory, only the *ab*-component of the current couples to the $\sigma$ bands, whereas the π bands are probed by currents in all directions. Therefore, if there were equal contributions to the DOS from the $\sigma$ and $\pi$ bands [21], we would still expect the geometrical factor to result in a higher measured weighting for the $\pi$ band. The effect is difficult to quantify, and so we have to rely on a best fit proceedure for obtaining the value of *f*. We have found that typically $f = 0.65 – 0.8$ in different contacts.

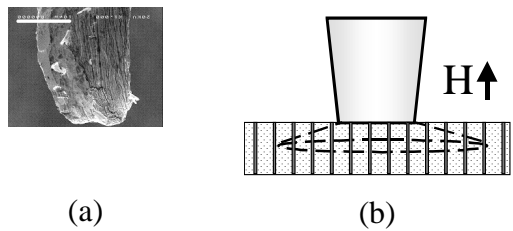

(a)          (b)

Fig 1. (a) Scanning electron microscopy image of the gold tip after the point-contact measurements. The scale bar corresponds to 50 microns. The size of the deformed tip apex is indicative of the contact dimensions. (b) Schematic representation of the contact probing a large number of Abrikosov vortices (vertical lines) with the current injected into a large solid angle (dashed lines).

In the presence of magnetic field, Eq (1) needs to be modified to take into account the effect of normal cores. We do so by introducing two further variables, $n_\pi$ and $n_\sigma$, to represent the field-dependent normal state contribution to the differential conductance in each of the two bands. The resulting formula is:

$$\frac{G(V)}{G_N} = f[n_\pi + (1-n_\pi)g_\pi] + (1-f)[n_\sigma + (1-n_\sigma)g_\sigma]$$

(2).

As $n_\pi$ and $n_\sigma$ represent the number of normal-state core excitations, these parameters can be identified with the zero-energy DOS $N_\pi^0$ and $N_\sigma^0$, where the latter are understood as values averaged over the vortex lattice.

Because of the large number of parameters involved, fitting the experimental data to Eq.(2) may yield non-unique results. To avoid this, we constrain some parameters to their zero-field values, as explained below.

The experiment in $H\|c$ corresponds to the situation considered theoretically in [4]. The data and respective best fits to Eq (2) are shown in Fig 2(a). The fitting is performed in two steps: At zero field the fit to Eq (1) is robust and we obtain the values of *Z* and *f*. Assuming that the interface properties (*Z*) and the injection geometry (*f*) remain the same unless the tip moves and the contact geometry changes, we treat these parameters as field-independent. In this case, adding two more variables in Eq (2) ($n_\pi$ and $n_\sigma$) for the in-field data does not compromise the robustness of the fit, so that the set of five parameters ($\Delta_\pi$, $\Delta_\sigma$, $\omega$, $n_\pi$ and $n_\sigma$) can be extracted reliably.

The field dependences of the order parameters, $\Delta_\pi$, $\Delta_\sigma$ obtained from the present analysis are consistent with [15]. In particular, we observe that $\Delta_\pi$ does not collapse at a field much smaller than the global $H_{c2}$. The effective broadening increases linearly with field. However the key result here is a very strong disparity between the field dependences of $n_\pi$ and $n_\sigma$.

As Fig 3 shows, $n_\pi$ increases towards its normal-state value of 1 significantly faster than $n_\sigma$. Within the field range where it was possible to obtain reliable fits, both parameters exhibit closely linear increase with *H*. The gradient of the $n_\sigma(H)$ dependence is close to $1/H_{c2}$, similar to both calculated and observed behaviour for a conventional superconductor such as Nb [18]. In contrast, $n_\pi(H)$ has a gradient at low field that is four times as steep. These results are in perfect qualitative agreement with the theory [4]. Quantitatively, the values of the gradients are controlled by the ratio of electron diffusivities in the two bands, $D_\sigma^{(ab)}/D_\pi^{(ab)}$ (the in-plane diffusivities are relevant in this field orientation). It is the value of $dn_\pi/dH$ that is very sensitive to the variation of $D_\sigma^{(ab)}/D_\pi^{(ab)}$, with the gradient being steep for a relatively dirty $\sigma$ band/clean $\pi$ band ($D_\sigma^{(ab)}/D_\pi^{(ab)} < 1$). We show the results of theoretical calculations [22] for $D_\sigma^{(ab)}/D_\pi^{(ab)} = 0.5$ in Fig 3. The conclusion is that the results of our experimental analysis in conjunction with the two-band dirty-limit theory can be used for accurate determination of the diffusivities ratio.

In field applied parallel to the ab plane of the sample, $H_{c2}$ at T=4.2 K is beyond the reach of our magnet; we therefore extrapolate the slowly varying anisotropy factor from higher temperatures and evaluate $H_{c2\|ab} \sim 30$ T. It is remarkable that $n_\pi(H)$ does not scale with $H_{c2\|ab}$, but rather follows the same dependence for the two field orientations. On the contrary, the behaviour of the $\sigma$ band normal state contribution does depend on the field direction. For $H\|ab$ we do not see any variation in $n_\sigma$ beyond the error margin up to 3 T, which is consistent with the fact that this field range corresponds to only 10 % of $H_{c2}$. Because the $\pi$ band diffusivity is isotropic, the anisotropy of $H_{c2}$ must arise from the difference between $D_\sigma^{(ab)}$ and $D_\sigma^{(c)}$. Indeed in both field orientations, the $\sigma$ band is the dirtier one and so it always determines $H_{c2}$ in our sample. Consequently [3,4], $D_\sigma^{(ab)}/D_\sigma^{(c)} \sim (H_{c2}^c/H_{c2}^{ab})^2 \sim 25$.

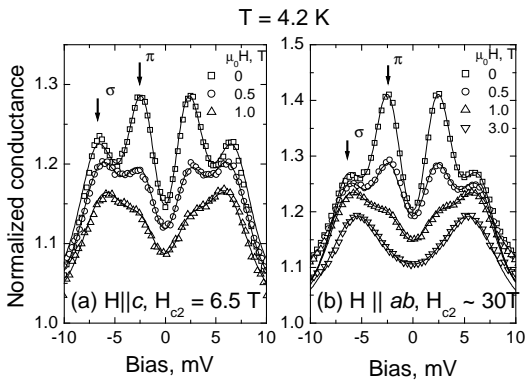

Fig 2. Representative experimental data at 4.2 K and corresponding best fits to Eq 2: (a) field parallel to the *c* axis; (b) field parallel to the *ab* plane.

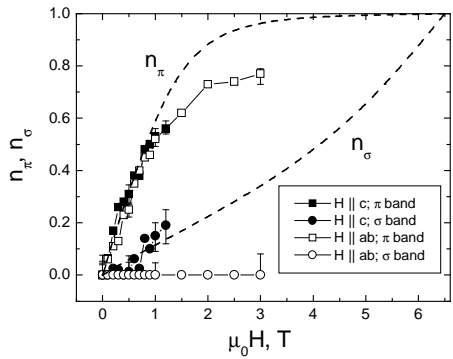

Fig. 3. Field variation of the normal state contribution to the DOS in the $\sigma$ and $\pi$ bands as a function of field, as inferred from the PCAR data using Eq 2. Solid symbols: $H\|c$; open symbols: $H\|ab$; circles: $\sigma$ band; squares: $\pi$ band. Dashed lines are theoretical results [22] for $D_\sigma^{(ab)}/D_\pi^{(ab)} = 0.5$ and $H_{c2\|c} = 6.5$ T.

The theoretical calculation of the vortex core structure in a two-band superconductor [4] strictly applies to the $H\|c$ case only. For the field applied parallel to the ab plane, the core is likely to be more complex due to the large disparity between the in-plane and out-of-plane $\sigma$-band Fermi velocities, and so may require a refined theory to correlate the experimental observables with the microscopic properties. As an indication of the possible complications, we note that only in the $H\|ab$ case have we observed that the parameter $f$ in (2) becomes field dependent, decreasing from 0.72 at zero field to 0.69 at 3 T.

In conclusion, we have demonstrated that the field variation of the Andreev reflection spectra in an MgB$_2$ film can be consistently understood in terms of the dirty-limit theory. In field applied parallel to the c axis, it is possible to determine the ratio of diffusivities quantitatively. This independent experimental protocol for evaluating the diffusivities may be particularly important for the analysis of the enhanced $H_{c2}$ in MgB$_2$ films [23], as the theory [3] requires these parameters for calculating $H_{c2}$ and its temperature dependence. The diffusivities are also key to understanding the anisotropy of $H_{c2}$ and the behaviour of the London penetration depth, and hence the microwave properties of MgB$_2$.

We thank A.Gurevich, I.I.Mazin and A.A.Golubov for useful discussions. The work at Penn State is supported in part by ONR under grant No. N00014-00-1-0294 and by NSF under grant No. DMR-0306746. This work was supported by the UK Engineering and Physical Sciences Research Council.

**Figure captions**

Fig 1. (a) Scanning electron microscopy image of the gold tip after the point-contact measurements. The scale bar corresponds to 50 microns. The size of the deformed tip apex is indicative of the contact dimensions. (b) Schematic representation of the contact probing a large number of Abrikosov vortices (vertical lines) with the current injected into a large solid angle (dashed lines).

Fig 2. Representative experimental data at 4.2 K and corresponding best fits to Eq 2: (a) field parallel to the *c* axis; (b) field parallel to the *ab* plane.

Fig. 3. Field variation of the normal state contribution to the DOS in the $\sigma$ and $\pi$ bands as a function of field, as inferred from the PCAR data using Eq 2. Solid symbols: $H\|c$; open symbols: $H\|ab$; circles: $\sigma$ band; squares: $\pi$ band. Dashed lines are theoretical results [22] for $D_\sigma^{(ab)}/D_\pi^{(ab)} = 0.5$ and $H_{c2\|c} = 6.5$ T.